# Odd non-linear conductivity under spatial inversion in chiral Tellurium

Manuel Suárez-Rodríguez[1], Beatriz Martín-García[1,2], Witold Skowroński[1,3], F. Calavalle[1], Stepan S. Tsirkin[2,4], Ivo Souza[2,4], Fernando De Juan[2,5], Andrey Chuvilin[1,2], Albert Fert[5,6,7], Marco Gobbi[2,4,\*], Fèlix Casanova[1,2,†], and Luis E. Hueso[1,2,‡].

[1]CIC nanoGUNE BRTA, 20018 Donostia-San Sebastián, Basque Country, Spain
[2]IKERBASQUE, Basque Foundation for Science, 48009 Bilbao, Basque Country, Spain
[3]AGH University of Science and Technology, Institute of Electronics, 30-059 Kraków, Poland
[4]Centro de Física de Materiales CSIC-UPV/EHU, 20018 Donostia-San Sebastián, Basque Country, Spain
[5]Donostia International Physics Center, 20018 Donostia-San Sebastián, Basque Country, Spain
[6]Unité Mixte de Physique, CNRS, Thales, Université Paris-Saclay, 91767 Palaiseau, France
[7]Department of Materials Physics UPV/EHU, 20018 Donostia-San Sebastián, Basque Country, Spain

Correspondence to: *marco.gobbi@ehu.eus; †f.casanova@nanogune.eu; ‡l.hueso@nanogune.eu

Electrical transport in non-centrosymmetric materials departs from the well-established phenomenological Ohm's law. Instead of a linear relation between current and electric field, a non-linear conductivity emerges along specific crystallographic directions. This non-linear transport is fundamentally related to the lack of spatial inversion symmetry. However, the experimental implications of an inversion symmetry operation on the non-linear conductivity remain to be explored. Here, we report on a large, non-linear conductivity in chiral Tellurium. By measuring samples with opposite handedness, we demonstrate that the non-linear transport is odd under spatial inversion. Furthermore, by applying an electrostatic gate, we modulate the non-linear output by a factor of 300, reaching the highest reported value excluding engineered heterostructures. Our results establish chiral Te as an ideal compound not just to study the fundamental interplay between crystal structure, symmetry operations and non-linear transport; but also to develop wireless rectifiers and energy-harvesting chiral devices.

*Introduction.*–Ohm's law is one of the most established relations in Physics and forms the basis of today's electronics. This phenomenological relation states that the charge current density ($j$) in a material is linearly proportional to the applied electric field ($E$) [1]. Under time-reversal symmetry conditions, most conducting materials follow this correlation down to the nanoscale [2], with a constant of proportionality which is the electrical conductivity of the system ($\sigma$). However, and starting from recent seminal studies in WTe$_2$ [3-6], different experimental [7-15] and theoretical [16-22] reports showed that, in systems with broken inversion symmetry, a non-linear charge current quadratically proportional to the applied electric field is allowed, $\boldsymbol{j} = \sigma\boldsymbol{E} + \chi\boldsymbol{E}\boldsymbol{E}$ [23]. In this case, the proportionality is captured by the non-linear susceptibility tensor of the system ($\chi$), which depends on the symmetry of the material and may present both diagonal and off-diagonal components. On the one hand, off-diagonal components, which result on transverse responses leading to the non-linear Hall effect, have attracted much attention as they are connected to the Berry curvature dipole [18]. On the other hand, fewer studies have been devoted to diagonal components, which are associated to the non-linear longitudinal conductivity [24] and are purely connected to extrinsic mechanisms [18, 25]. These components could lead to strong non-linear output signals promising for applications in frequency-doubling [26, 27], energy harvesting and wireless detection via rectification [28-30].

Since breaking inversion symmetry is the fundamental requirement for the emergence of non-linear transport, an inversion symmetry operation (*i*) is expected to profoundly affect the non-linear conductivity. However, the implications of such operation have not been experimentally explored so far. Indeed, the symmetry of the studied systems, mainly restricted to few-layer polar transition-metal dichalcogenides [31-33] and engineered non-centrosymmetric structures [34-40], does not permit to distinguish the effect of *i* from that of a simple rotation.

Chiral materials have neither mirror planes nor inversion centers, and are, therefore, non-centrosymmetric. Unlike trivial rotations, *i* results in a change of handedness. In spite of their importance to test the effect of *i*, non-linear conductivity has not been explored in any chiral material. In this regard, chiral elemental Te, which can be synthesized on single crystalline flakes with both handedness [41-44] and excellent electronic properties [45], represent an ideal platform to disentangle the effect of *i* and rotation on the non-linear transport.

Here, we report the observation of a large non-linear conductivity in chiral single crystalline Te flakes. The dependence of this non-linear effect on the relative orientation between the crystallographic axes of Te and the current direction is fully explained by symmetry considerations, as it arises from the non-linear susceptibility tensor of the space group of Te. By performing measurements on Te flakes with opposite handedness, we experimentally demonstrate that *i* changes the sign of the non-linear tensor, in agreement with its spatial-inversion-odd nature. Moreover, the resistivity dependence of the non-linear conductivity shows that side-jump scattering is its dominant microscopic mechanism. Finally, the electrostatic gating of the Te flakes allows us to tune its non-linear response by a factor of 300, reaching a non-linear voltage of up to 1.2 mV. This strong output signal, the highest reported excluding engineered heterostructures, opens the path to exploit chiral materials for wireless rectification and energy harvesting applications.

*Anisotropic transport.*–Elemental Te is the simplest material with a chiral structure. It is constituted by covalently-bonded Te helices along the *z*-axis, stacked together by van der Waals interactions. Single-crystalline Te flakes were grown by a hydrothermal process in the presence of a reducing agent. Tens-of-micrometers-long, few-micrometers-wide, and a hundred-nanometer-thick flakes were transferred onto Si/SiO$_2$ substrates using a Langmuir-Schaefer approach. We note that the flakes lie on the substrate with their *x*- and chiral *z*-axis in plane. In order to control the current direction and enable precise electrical measurements along the different crystal axes, the flakes were patterned in a star-like shape by reactive-ion etching, and contacted by sputtered Pt (see Supplemental Section 1 [46]). Figure 1(a) shows the scanning electron microscopy (SEM) image of a typical device used in this study, with a sketch of the measurement configuration. A harmonic current, $I^\omega$ (31 Hz) is injected between two electrodes which are along the same direction at an angle $\theta$ from the chiral *z*-axis, while both the longitudinal ($V_\parallel$) and transverse ($V_\perp$) voltages are measured in a rotating reference frame, meaning that all probes rotate simultaneously.

The temperature dependence of the first-harmonic longitudinal resistance along the crystal *x*-axis ($R_{xx}^\omega$) and *z*-axis ($R_{zz}^\omega$) is plotted in Fig. 1(b). The resistivity along the *x*-axis is ~10 times higher than along the *z*-axis at room temperature, and it increases much more rapidly with decreasing temperature. This is a consequence of the anisotropic crystal structure of Te, with more favorable electronic transport along the covalently-bonded Te helices, which are aligned along the chiral *z*-axis. The first-harmonic voltage scales linearly with the current (Supplemental Section 2 [46]). The slopes corresponding to the first-harmonic longitudinal $\left(R_\parallel^\omega \equiv \frac{V_\parallel^\omega}{I^\omega}\right)$ and transverse $\left(R_\perp^\omega \equiv \frac{V_\perp^\omega}{I^\omega}\right)$ resistances as a function of $\theta$ at 10 K are presented in Fig. 1(c). Both show a two-fold angular dependence, which is consistent with the crystal symmetry of Te and can be expressed as $R_\parallel^\omega(\theta) = R_{zz}^\omega \cos^2\theta + R_{xx}^\omega \sin^2\theta$ and $R_\perp^\omega(\theta) = (R_{xx}^\omega - R_{zz}^\omega)\cos\theta\sin\theta$. The fit [dashed lines, Fig. 1(c)] yields a resistance anisotropy $r \left(\equiv \frac{R_{xx}}{R_{zz}}\right)$ of about 140. All measured devices display similar behavior (Supplemental Section 3 [46]).

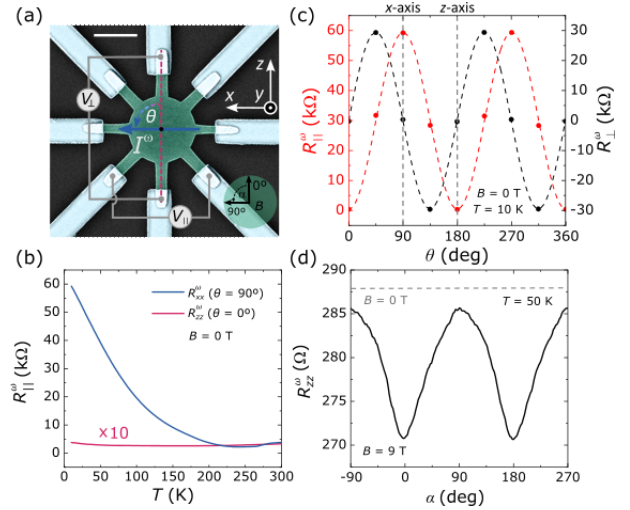

FIG. 1. Basic characterization of Te. (a) False-color SEM image of a typical star-like device used in this study. The etched Te flake (green) is contacted by Pt contacts (blue). The scale bar corresponds to 2 μm. (b) Temperature dependence of the parallel first-harmonic resistance ($R_\parallel^\omega$) when $I^\omega$ is injected along *x*- and *z*-axis. The data points for the latter is multiplied by 10 for clarity. (c) Parallel ($R_\parallel^\omega$) and transverse ($R_\perp^\omega$) first-harmonic resistance as a function of $\theta$ at 10 K. The dashed lines are fits to the equations introduced in the main text. (d) $\alpha$-angle-dependent magnetoresistance of $R_{zz}^\omega$ at 9 T and 50 K. The grey dashed line indicates the $R_{zz}^\omega$ value at zero field. All data were measured in device S1.

Fig. 1(d) shows the angular-dependent magnetoresistance of $R_{zz}^\omega$ obtained at 50 K by rotating an external magnetic field ($B$ = 9 T) an in-plane angle $\alpha$. The curves are characterized by minima at around $\alpha = 0°$ and $\alpha = 180°$, that is, for $B \parallel I_z$, and maxima at around $\alpha = \pm 90°$, for $B \perp I_z$. We note that the resistance measured at $B = 0$ T, which is shown as a dashed line, is higher for all $\alpha$ angles, indicating that the studied Te flakes present negative magnetoresistance for any in-plane direction of $B$, but stronger when it is aligned along the $z$-axis.

*Non-linear conductivity.*–The chiral structure of Te allows the generation of a second-order current density when an electric field is applied, $\boldsymbol{j} = \sigma\boldsymbol{E} + \chi\boldsymbol{EE}$. That is, the non-linear susceptibility tensor for the space groups of Te, $P3_221$ (left-handed) or $P3_121$ (right-handed) present non-vanishing elements [48]:

$$\chi = \begin{pmatrix} \chi_{xxx} & \chi_{xyy} & 0 & \chi_{xyz} & 0 & 0 \\ 0 & 0 & 0 & 0 & \chi_{yxz} & \chi_{yxy} \\ 0 & 0 & 0 & 0 & 0 & 0 \end{pmatrix} \quad (1)$$

where $\chi_{xxx} = -\chi_{xyy} = -\chi_{yxy}$ and $\chi_{xyz} = -\chi_{yxz}$. In our experimental set-up, we measure the components in the $x$-$z$ plane [Fig. 1(a)]. Hence, we are sensitive to the diagonal component $\chi_{xxx}$, which acts as a correction to the Ohm's law along the $x$-axis, $j_x = \sigma_{xx}E_x^\omega + \chi_{xxx}E_x^\omega E_x^\omega$. Therefore, when an electric field ($E^\omega$) is applied along the $x$-axis, a non-linear current density is generated in the same direction [Fig. S3(a)]. Conversely, when an electric field is applied along the $z$-axis, the generation of a non-linear charge current is forbidden by symmetry constrains, and therefore, Ohm's law holds, $j_z = \sigma_{zz}E_z^\omega$ [Fig. S3(b)].

For an in-plane a.c. current at an angle $\theta$ from the chiral $z$-axis, the non-linear response through the second-order non-linear susceptibilities can be expressed as follows (see Supplemental Section 4 [46]):

$$\frac{V_\parallel^{2\omega}}{(I^\omega)^2} = \frac{L_\parallel}{L_x^3} A(R_{xx}^\omega)^3 \chi_{xxx} \sin^3(\theta) \quad (2)$$

$$\frac{V_\perp^{2\omega}}{(I^\omega)^2} = \frac{L_\perp}{L_x^3} A(R_{xx}^\omega)^3 \chi_{xxx} \sin^2(\theta)\cos(\theta) \quad (3)$$

where $V_\parallel^{2\omega}$ and $V_\perp^{2\omega}$ are the second-harmonic voltages generated in a direction longitudinal and transverse to the current direction; $L_\parallel$, $L_\perp$, and $L_x$ are the distances between contacts along the parallel direction, perpendicular direction, and $x$-axis, respectively; $A$ is the cross section. The equations take into account that, experimentally, we apply currents and measure voltages.

We highlight that we employ both a longitudinal and a transverse configuration to probe the diagonal component $\chi_{xxx}$, which describes the longitudinal non-reciprocal transport along the $x$-axis. This is a unique feature that is fundamentally related to Te symmetry and proves the breakdown of Ohm's law. Moreover, the presence of just one component in equations (2) and (3) allows us to fully determine its value. Previous studies employed the transverse (Hall) configuration to probe off-diagonal [4,8,9], but also a combination of diagonal and off-diagonal components [5] of the non-linear tensor. For this reason, the non-linear transport is often referred to as non-linear Hall effect.

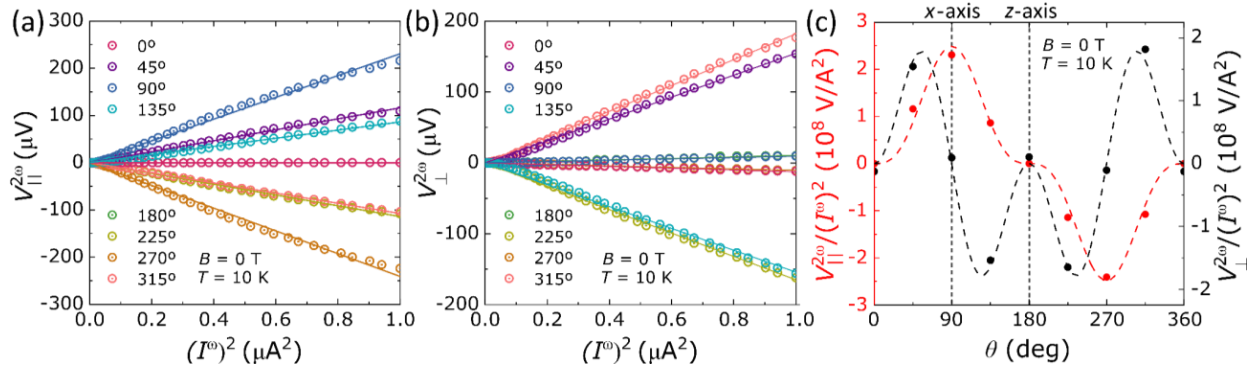

FIG. 2. Non-linear conductivity. (a),(b) Second-order parallel $V_\parallel^{2\omega}$ (a) and transverse $V_\perp^{2\omega}$ (b) voltage as a function of $(I^\omega)^2$ at different $\theta$ angles. The second-order voltage depends linearly on the square of $I^\omega$. (c) The non-linear conductivity [slopes of the dependences in (a) and (b)] as a function of $\theta$. The dashed lines are fits to the experimental data with equations (2) and (3). All measurements were performed in device S1 at 10 K.

We biased the disc device S1 with a harmonic current in a sequence along the 8 contacts rotated by 45 degrees. The longitudinal, $V_\parallel^{2\omega}$ [Fig. 2(a)] and transverse, $V_\perp^{2\omega}$ [Fig. 2(b)] second-harmonic voltages were recorded in a rotating reference frame. Both scale linearly with the square of the current, and switch sign when the current direction and the voltage probe connections are reversed simultaneously. The slopes of $V_\parallel^{2\omega}$ and $V_\perp^{2\omega}$ normalized to $(I^\omega)^2$ as a function of $\theta$ are summarized in Fig. 2(c). Equations (2) and (3) perfectly capture the experimental angular dependence (dashed lines). From the two fittings, we obtain similar values of the non-linear susceptibility component: $\chi_{xxx} = (2.13 \pm 0.18) \times 10^{-3}$ and $(2.49 \pm 0.18) \times 10^{-3}$ $\Omega^{-1}\cdot V^{-1}$ from the longitudinal and transverse signals, respectively. Therefore, the experimental response is consistent with a non-linear effect based on the symmetry of chiral Te.

*Spatial inversion operation.*– The operation of spatial inversion ($i$), which is the combination of a 2-fold rotation with a perpendicular reflection, is fundamentally different from just a trivial rotation when applied to a chiral structure, as it introduces a change of handedness. Therefore, the Te flakes, which are synthesized as single crystals with both left and right handedness, allow us to directly explore the effect of $i$ on the non-linear conductivity [Fig. 3(a)]. We must point out that $i$ in Te does not only change its handedness as it also changes the orientation of the atomic triangles pattern that Te displays in the *x-y* plane. These triangular units can be imaged by scanning transmission electron microscopy (STEM), as shown in Fig. 3(a). Therefore, for comparing samples with different handedness and test the effect of $i$, we must align them with their atomic triangular pattern pointing to +$x$ and –$x$, respectively.

We present the non-linear conductivity in two Te devices with opposite handedness (device S1 and S2). Their handedness was determined by measuring the unidirectional magnetoresistance (UMR), also known as electrical magnetochiral anisotropy [43]. The second-harmonic resistance along z-axis, $R_{zz}^{2\omega}$, as a function of $\alpha$-angle at 9 T gives us the UMR, showing a peak (valley) at $\alpha = 0$ ($\alpha = 180°$) for device S1 [Fig. 3(b)] and a valley (peak) at $\alpha = 0°$ ($\alpha = 180°$) for device S2 [Fig. 3(c)]. The UMR in Te is a consequence of a chirality-dependent Edelstein effect and confirms that devices S1 and S2 are left- and right-handed, respectively (see Supplemental Section 5 [46]).

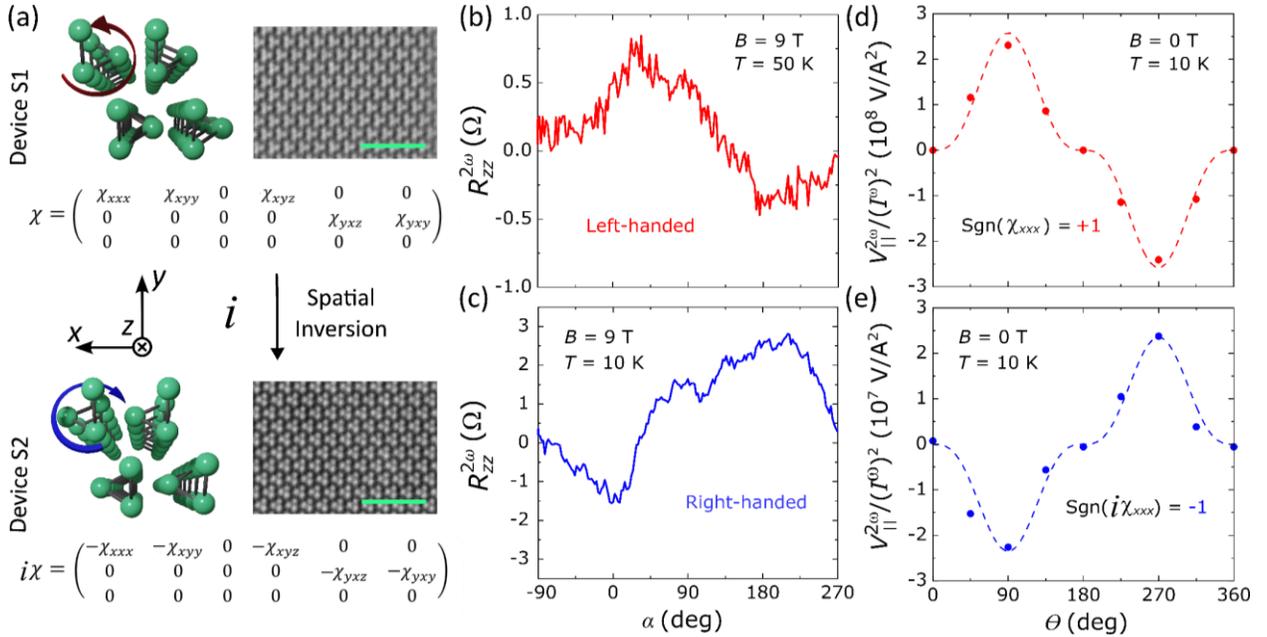

FIG. 3. Spatial inversion operation. (a) Spatial inversion operation over trigonal Te results on a change of handedness, 180° rotation of the *x-y* plane triangles pattern, and sign change of the non-linear susceptibility tensor. STEM images of Te lamellas obtained by cutting perpendicular to the *z* axis devices S1 and S2 (scale bars correspond to 2 nm). (b),(c) $\alpha$-angle-dependent magnetoresistance of $R_{zz}^{2\omega}$ at 9T measured at 50 K in device S1 (b) and at 10 K in device S2 (c). The $R_{zz}^\omega$ signal determines that devices S1 and S2 are left- and right-handed, respectively. (d),(e) Non-linear conductivity as a function of $\theta$ at 10 K for device S1 (d) and S2 (e). The dashed lines are fits to the experimental data with equation (2). Data and fit in (d) are the same as in Fig. 2(c).

By imaging the Te x-y plane through STEM, we can orient the device S1 (left-handed) and S2 (right-handed) with their atomic triangular pattern pointing to +x and −x, respectively [STEM images in Fig. 3(a)]. We should highlight that the UMR measurement is independent on the orientation of x-axis, so the handedness of Te flakes is determined unambiguously [49].

The non-linear conductivity on devices S1 [Fig. 3(d)] and S2 [Fig. 3(e)] as a function of $\theta$ [slopes of the dependences in Fig. 2(a) and Fig. S6(a)] show for both the symmetry expected for Te [equation (2)], but opposite sign. Consequently, by measuring flakes with opposite handedness, we were able to demonstrate that the non-linear susceptibility ($\chi_{xxx}$) changes sign under spatial inversion, in agreement with its inversion-symmetry-odd nature. Therefore, the symmetry of chiral Te allowed us to directly probe the fundamental connection between non-linear conductivity and broken inversion symmetry.

*Temperature and electrostatic gate dependences.*–Finally, we studied the microscopic mechanism of the non-linear conductivity by observing its dependence on the resistivity of the material. The resistivity was modulated both by varying the sample temperature [Fig. 4(a)] and by applying an electrostatic gate voltage [Fig. 4(b)]. On the one hand, the resistivity increases while decreasing temperature, as also observed in Fig. 1(b). On the other hand, Te is a hole-doped semiconductor [45], therefore, its resistivity increases for positive gate voltages. For the same temperatures and gate voltages, we studied the non-linear transverse voltage at $\theta = 45°$ and $\theta = 135°$. As also observed in Fig. 3(b), it depends linearly on $(I^\omega)^2$. Its slope decreases while increasing temperature [Fig. S7(a)], and it increases by applying positive gate voltages [Fig. S7(b)]. The non-linear conductivity as a function of the resistivity is plotted in Fig. 4(c) and Fig. 4(d). The scaling law between these two magnitudes can be formulated as [16,50]:

$$\frac{V_\perp^{2\omega}}{(I^\omega)^2} = \beta + \gamma \rho_{xx} + \delta \rho_{xx}^2 \quad (4)$$

where $\delta$ is an independent parameter, and $\beta$ and $\gamma$ only depend on the residual resistivity of the material. In time-reversal invariant systems, the only intrinsic contribution to the non-linear susceptibility tensor is the Berry curvature dipole, which is captured by off-diagonal components [18,25]. Since we are probing $\chi_{xxx}$ (diagonal component), the measured effect cannot have an intrinsic origin. The fit to equation (4) [dashed lines in Fig. 4(c) and Fig. 4(d)] reveals a non-monotonic behavior of $V_\perp^{2\omega}/(I^\omega)^2$ versus $\rho_{xx}$, where $V_\perp^{2\omega}/(I^\omega)^2$ first decreases and then increases. This requires the linear and quadratic coefficients to have different signs [Fig. 4(d)]. The existence of a linear term in $\rho_{xx}$ with opposite sign can only be consistent with a mechanism involving side-jump scattering from dynamic sources (see detailed analysis in Supplemental Section 7 [46]). Moreover, the similar temperature- and gate-dependence of the observed non-linear conductivity evidences a negligible contribution from a displacement field when applying an electrostatic gate voltage. Therefore, the strong intrinsic low symmetry of chiral Te is not substantially affected by the presence of an out-of-plane electric field.

Besides the fundamental connection between the non-linear conductivity in chiral Te and the side-jump microscopic mechanism, the electrostatic gate allows us to modulate the second-harmonic voltage output by a factor of 300, reaching a value of 1.2 mV for 1 μA input current [Fig. 4(c)]. Although the non-linear susceptibility components would be the most appropriate quantities to compare between different systems, they are not reported in previous literature. Therefore, and because of its potential importance in energy harvesting applications, we highlight that our voltage output is the highest reported excluding engineered heterostructures [Fig. S9].

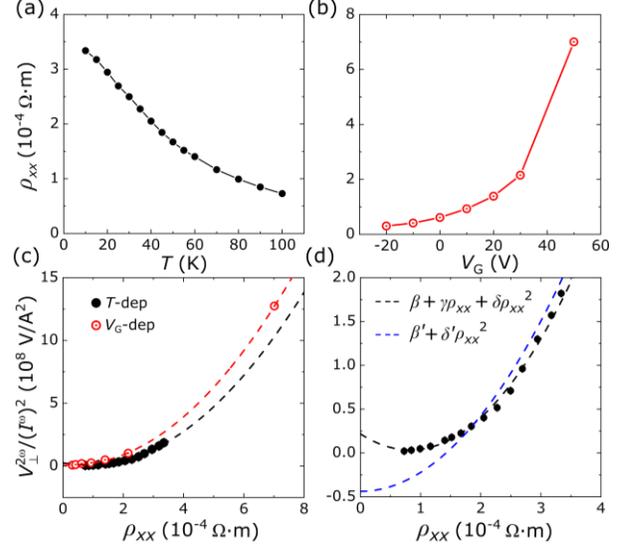

FIG. 4. (a) Temperature and (b) gate dependence of the resistivity along x-axis ($\rho_{xx}$). (c) The non-linear conductivity [taken from the slopes of $V_\perp^{2\omega}$ as a function of $(I^\omega)^2$ at $\theta = 45°$ in Figs. S7(a),(b)] as a function of $\rho_{xx}$, tuned by either temperature (solid black circles) or gate (open red circles). The dashed lines are fits to experimental data with equation (4). (d) Comparison between fits to equation (4) with (black dashed line) and without (blue dashed line) the linear term ($\gamma$) for the temperature-tuned dataset.

To conclude, we directly observed non-linear conductivity along the *x*-axis of chiral Te, which is odd under spatial inversion and the largest reported in a single material. Its resistivity dependence unveils side-jump scattering as the dominant microscopic mechanism.

*Acknowledgments.* – We thank Talieh S. Ghiasi for valuable insights regarding the device fabrication process. This work is supported by the Spanish MICINN (Project No. PID2021-122511OB-I00 and "Maria de Maeztu" Units of Excellent Programme No. CEX2020-001038-M). M.S.-R. acknowledges support from La Caixa Foundation (no. 100010434) with code LCF/BQ/DR21/11880030. B.M.-G. and M.G. thank support from the "Ramón y Cajal" Programme by the Spanish MCIN/AEI (grant no. RYC2021-034836-I and RYC2021-031705-I). W.S. thanks grant No. PPN/BEK/2020/1/00118/DEC/1 from Polish National Agency for Academic Exchange.

**References**

[1] D. J. Griffiths. Introduction to Electrodynamics. (Pearson, 1942).

[2] B. Weber et al. Ohm's law survives to the atomic scale. Science **335**, 64–67 (2012).

[3] Y. Zhang, J. van den Brink, C. Felser, B. Yan, Electrically tuneable nonlinear anomalous Hall effect in two-dimensional transition-metal dichalcogenides $WTe_2$ and $MoTe_2$. 2D Mater. **5**, 044001 (2018).

[4] Q. Ma et al. Observation of the nonlinear Hall effect under time-reversal-symmetric conditions. Nature **565**, 337–342 (2019).

[5] K. Kang, T. Li, E. Sohn, J. Shan, K. F. Mak, Nonlinear anomalous Hall effect in few-layer $WTe_2$. Nat. Mater. **18**, 324–328 (2019).

[6] J. Xiao et al. Berry curvature memory through electrically driven stacking transitions. Nat. Phys. **16**, 1028-1034 (2020).

[7] S. C. Ho et al. Hall effects in artificially corrugated bilayer graphene without breaking time-reversal symmetry. Nat. Electron. **4**, 116–125 (2021).

[8] A. Tiwari et al. Giant c-axis nonlinear anomalous Hall effect in $T_d$-$MoTe_2$ and $WTe_2$. Nat. Commun. **12**, 2049 (2021).

[9] T. Ma et al. Growth of bilayer $MoTe_2$ single crystals with strong nonlinear Hall effect. Nat. Commun. **13**, 5465 (2022).

[10] Y. M. Itahashi et al. Giant second harmonic transport under time-reversal symmetry in a trigonal superconductor. Nat. Commun. **13**, 1659 (2022).

[11] O. O. Shvetsov, V. D. Esin, A. V. Timonina, N. N. Kolesnikov, E. V. Deviatov, Nonlinear Hall Effect in Three-Dimensional Weyl and Dirac Semimetals. JETP Lett. **109**, 715–721 (2019).

[12] M.-S. Qin et al. Strain tunable Berry curvature dipole, orbital magnetization and nonlinear Hall effect in $WSe_2$ monolayer. Chinese Phys. Lett. **38**, 017301 (2021).

[13] S. Dzsaber et al. Giant spontaneous Hall effect in a nonmagnetic Weyl–Kondo semimetal. Proc. Natl. Acad. Sci. U. S. A. **118**, e2013386118 (2021).

[14] T. Nishijima et al. Ferroic Berry curvature dipole in a topological crystalline insulator at room temperature. Nano Lett. **23**, 2247-2252 (2023).

[15] L. Min et al. Colossal nonreciprocal Hall effect and broadband frequency mixing due to a room temperature nonlinear Hall effect. arXiv:2303.03738 [cond-mat.mes-hall] (2023).

[16] Z. Z. Du, C. M. Wang, S. Li, H. Z. Lu, X. C. Xie, Disorder-induced nonlinear Hall effect with time-reversal symmetry. Nat. Commun. **10**, 3047 (2019).

[17] Z. Z. Du, C. M. Wang, H. P. Sun, H. Z. Lu, X. C. Xie, Quantum theory of the nonlinear Hall effect. Nat. Commun. **12**, 5038 (2021).

[18] I. Sodemann, L. Fu, Quantum nonlinear Hall effect induced by Berry curvature dipole in time-reversal invariant materials. Phys. Rev. Lett. **115**, 216806 (2015).

[19] S. Nandy, I. Sodemann, Symmetry and quantum kinetics of the nonlinear Hall effect. Phys. Rev. B **100**, 195117 (2019).

[20] R. H. Li, O. G. Heinonen, A. A. Burkov, S. S.-L. Zhang, Nonlinear Hall effect in Weyl semimetals induced by chiral anomaly. Phys. Rev. B **103**, 045105 (2021).

[21] C. Wang, Y. Gao, D. Xiao, Intrinsic nonlinear Hall effect in antiferromagnetic tetragonal CuMnAs. Phys. Rev. Lett. **127**, 277201 (2021).

[22] H. Liu et al. Intrinsic second-order anomalous Hall effect and its application in compensated antiferromagnets. Phys. Rev. Lett. **127**, 277202 (2021).

[23] Z. Z. Du, H.-Z. Lu, X. C. Xie. Nonlinear Hall effects. Nat. Rev. Phys. **3**, 744-752 (2021).

[24] P. He et al. Graphene moiré superlattices with giant quantum nonlinearity of chiral Bloch electrons. Nat. Nanotechnol. **17**, 378–383 (2022).

[25] S. S. Tsirkin, I. Souza, On the separation of Hall and Ohmic nonlinear responses. SciPost Phys. Core **5**, 039 (2022).

[26] P. He et al. Quantum frequency doubling in the topological insulator $Bi_2Se_3$. Nat. Commun. **12**, 698 (2021).

[27] L. Min et al. Strong room-temperature bulk nonlinear Hall effect in a spin-valley locked Dirac material. Nat. Commun. **14**, 364 (2023).

[28] D. Kumar et al. Room-temperature nonlinear Hall effect and wireless radiofrequency rectification in Weyl semimetal $TaIrTe_4$. Nat. Nanotechnol. **16**, 421-425 (2021).

[29] Y. Zhang, L. Fu. Terahertz detection based on nonlinear Hall effect without magnetic field. Proc. Natl. Acad. Sci. U. S. A. **118**, e2100736118 (2021).

[30] C. Guo et al. Ultrasensitive anisotropic room-temperature terahertz photodetector based on an intrinsic magnetic topological insulator $MnBi_2Te_4$. Nano Lett. **22**, 7492–7498 (2022).

[31] B. T. Zhou, C.-P. Zhang, K. T. Law, Highly tunable nonlinear Hall effects induced by spin-orbit couplings in strained polar transition-metal dichalcogenides. Phys. Rev. Appl. **13**, 024053 (2020).

[32] S. Lai et al. Third-order nonlinear Hall effect induced by the Berry-connection polarizability tensor. Nat. Nanotechnol. **16**, 869–873 (2021).

[33] X.-G. Ye et al. Control over Berry curvature dipole with electric field in $WTe_2$. Phys. Rev. Lett. **130**, 016301 (2023).

[34] L. Du et al. Engineering symmetry breaking in 2D layered materials. Nat. Rev. Phys. **3**, 193–206 (2021).

[35] M. Huang et al. Giant nonlinear Hall effect in twisted bilayer WSe2. Natl. Sci. Rev. doi:10.1093/nsr/nwac232 (2022).

[36] J. Duan et al. Giant second-order nonlinear Hall effect in twisted bilayer graphene. Phys. Rev. Lett. **129**, 186801 (2022).

[37] S. Sinha et al. Berry curvature dipole senses topological transition in a moiré superlattice. Nat. Phys. **18**, 765-770 (2022).

[38] K. Kang et al. Switchable moiré potentials in ferroelectric WTe2/WSe2 superlattices. Nat. Nanotechnol. **18**, 861-866 (2023).

[39] M. Huang et al. Intrinsic nonlinear Hall effect and gate-switchable Berry curvature sliding in twisted bilayer graphene. Phys. Rev. Lett. **131**, 066301 (2023)

[40] N. J. Zhang et al. Angle-resolved transport nonreciprocity and spontaneous symmetry breaking in twisted trilayer graphene. arXiv:2209.12964 [cond-mat.mes-hall] (2023).

[41] A. Ben-Moshe et al. The chain of chirality transfer in tellurium nanocrystals. Science **372**, 729–733 (2021).

[42] Z. Dong, Y. Ma, Atomic-level handedness determination of chiral crystals using aberration-corrected scanning transmission electron microscopy. Nat. Commun. **11**, 1588 (2020).

[43] F. Calavalle et al. Gate-tuneable and chirality-dependent charge-to-spin conversion in tellurium nanowires. Nat. Mater. **21**, 526–532 (2022).

[44] A. Roy et al. Long-range current-induced spin accumulation in chiral crystals. npj Comput. Mater. **8**, 243 (2022)

[45] Y. Wang et al. Field-effect transistors made from solution-grown two-dimensional tellurene. Nat. Electron. **1**, 228–236 (2018).

[46] See Supplemental Material at [URL] for methods details, further experimental data and detailed theoretical analysis, which includes Ref. [47].

[47] P. He et al. Bilinear magnetoelectric resistance as a probe of three-dimensional spin texture in topological surface states. Nat. Phys. **14**, 495–499 (2018).

[48] S. V. Gallego, J. Etxebarria, L. Elcoro, E. S. Tasci, J. M. Perez-Mato, Automatic calculation of symmetry-adapted tensors in magnetic and non-magnetic materials: a new tool of the Bilbao Crystallographic Server. Acta. Cryst. **75**, 438–447 (2019).

[49] X. Liu, I. Souza, S. S. Tsirkin, Electrical magnetochiral anisotropy in trigonal tellurium from first principles. arXiv:2303.10164 [cond-mat.mtrl-sci] (2023).

[50] D. Hou et al. Multivariable scaling for the anomalous Hall effect. Phys. Rev. Lett. **114**, 217203 (2015).

**Supplemental material for**

# Odd non-linear conductivity under spatial inversion in chiral Tellurium

Manuel Suárez-Rodríguez[1], Beatriz Martín-García[1,2], Witold Skowroński[1,3], F. Calavalle[1], Stepan S. Tsirkin[2,4], Ivo Souza[2,4], Fernando De Juan[2,5], Andrey Chuvilin[1,2], Albert Fert[5,6,7], Marco Gobbi[2,4,*], Fèlix Casanova[1,2,†], and Luis E. Hueso[1,2,‡].

[1]*CIC nanoGUNE BRTA, 20018 Donostia-San Sebastián, Basque Country, Spain*
[2]*IKERBASQUE, Basque Foundation for Science, 48009 Bilbao, Basque Country, Spain*
[3]*AGH University of Science and Technology, Institute of Electronics, 30-059 Kraków, Poland*
[4]*Centro de Física de Materiales CSIC-UPV/EHU, 20018 Donostia-San Sebastián, Basque Country, Spain*
[5]*Donostia International Physics Center, 20018 Donostia-San Sebastián, Basque Country, Spain*
[6]*Unité Mixte de Physique, CNRS, Thales, Université Paris-Saclay, 91767 Palaiseau, France*
[7]*Department of Materials Physics UPV/EHU, 20018 Donostia-San Sebastián, Basque Country, Spain*

Correspondence to: *marco.gobbi@ehu.eus; †f.casanova@nanogune.eu;
‡l.hueso@nanogune.eu

**Contents**                                                                                                                       Pg.

## Section 1. Methods

*I. Chemical Synthesis of Te flakes.* – As in our previous paper [43], we synthesized the Te flakes by chemically reducing tellurium oxide in presence of hydrazine ($N_2H_4$) in a basic aqueous medium at high temperature. Here, we carried out slight modifications in the synthesis. Briefly, we dissolved at room temperature $Na_2TeO_3$ (104 mg) and polyvinylpyrrolidone (average $M_w$ 29,000 – PVP 29, 92.1 mg) in 33 mL of type I water by magnetic stirring. Then, $NH_4OH$ solution (3.65 mL, 25%w in water) and hydrazine hydrate (1.94 mL, 80%, w/w%) were added while stirring. The mixture was transferred to an autoclave that was sealed and heated at 180°C for 23 h. We purified the resulting material by successive centrifuge-assisted precipitation (10,000 rpm – 1h, Avanti J-26 XPI centrifuge) and redispersion with type I water. Finally, the Te flakes were redispersed in a dimethylformamide/$CHCl_3$ mixture (1.3:1 v/v) to be used in the sample fabrication. All the reagents were purchased from Sigma Aldrich and used as received without any further purification.

*II. Sample fabrication.* – Te flakes were redispersed in a dimethylformamide/$CHCl_3$ mixture (1.3:1 v/v) to be used for the drop-casting of solution droplets at the type I-water/air interface in a homemade Langmuir trough. After the evaporation of the solvent, Te flakes floating on the water surface were picked up (Langmuir–Schaefer technique) with Si/$SiO_2$ substrates (Si doped n+, 5 × 5 mm, 300 nm thermal oxide). Isolated flakes with suitable dimensions were selected with an optical microscope, without knowing a priori the handedness of the flakes. The flake shape and the contacts were defined through electron-beam lithography performed on a poly(methylmethacrylate)-A4/poly(methylmethacrylate)-A2 double layer. The patterning of the flakes was performed by reactive ion etching (Ar gas, 50 W RF power, 5 x 2 min), and Pt was deposited by sputtering for the electrical contacts. Te has a very anisotropic electrical resistivity and, therefore, the patterning is needed to properly control the current directionality. Six flake-based devices (S1–S6) were fabricated and measured at low temperatures.

*III. Electrical measurements.* – The devices were wire-bonded to a sample holder and installed in a physical property measurement system (Quantum Design) for transport measurements with a temperature range of 2–400 K and maximum magnetic field of 9 T. We performed measurements of a.c and d.c. longitudinal and transverse resistance, using a Keithley 6221 current source (with current ranging from 50 nA to 1 µA). We keep our maximum current at 1 µA to reduce heating and obtain a uniform resistivity value over the entire current range. On the one hand, for the a.c. measurements, the longitudinal and transverse voltage drops were measured at the fundamental and the second-harmonic frequencies with a dual channel NF LI5660 lock-in amplifier. Fundamental frequency was set to $\omega$ = 31 Hz. On the other hand, for the magnetotransport measurements, we record the d.c. longitudinal voltage with a Keithley 2128 nanovoltmeter. Asymmetric ($I^+$/0 and $I^-$/0) delta mode (16 to 32 counts) was employed to improve the signal-to-noise ratio (with a current of 1 µA). The average of the signals obtained from the asymmetric delta mode measurements corresponds to the first-harmonic resistance in a.c. measurements ($R^\omega$). The half difference of the signals obtained from the asymmetric delta mode was taken to obtain the equivalent second-harmonic resistance ($R^{2\omega}$) [47]. For the gate-dependence measurements, a Keithley 2636 source meter was used to apply a constant voltage to doped Si substrate, while monitoring the leakage current through the SiO$_2$ dielectric to be smaller than 2 nA. We note that for the longitudinal measurements, we obtained the signal at both sides of the flake, and we presented the averaged value.

*IV. Transmission electron microscopy imaging.*–The lamellae were fabricated from tested devices perpendicular to *z*-axis by a standard FIB lift-out technique (Helios 450S FIB/SEM, Thermo Fisher Scientific, USA). In order to preserve the orientation upon potential lamellae swapping while mounting in TEM holder one side of the lamellae was marked by a cut visible in TEM. The lamellae were studied in Titan 60-300 (S)TEM (Thermo Fisher Scientific, USA) in STEM mode at 300 kV.

## Section 2. Extended data of anisotropic transport on device S1.

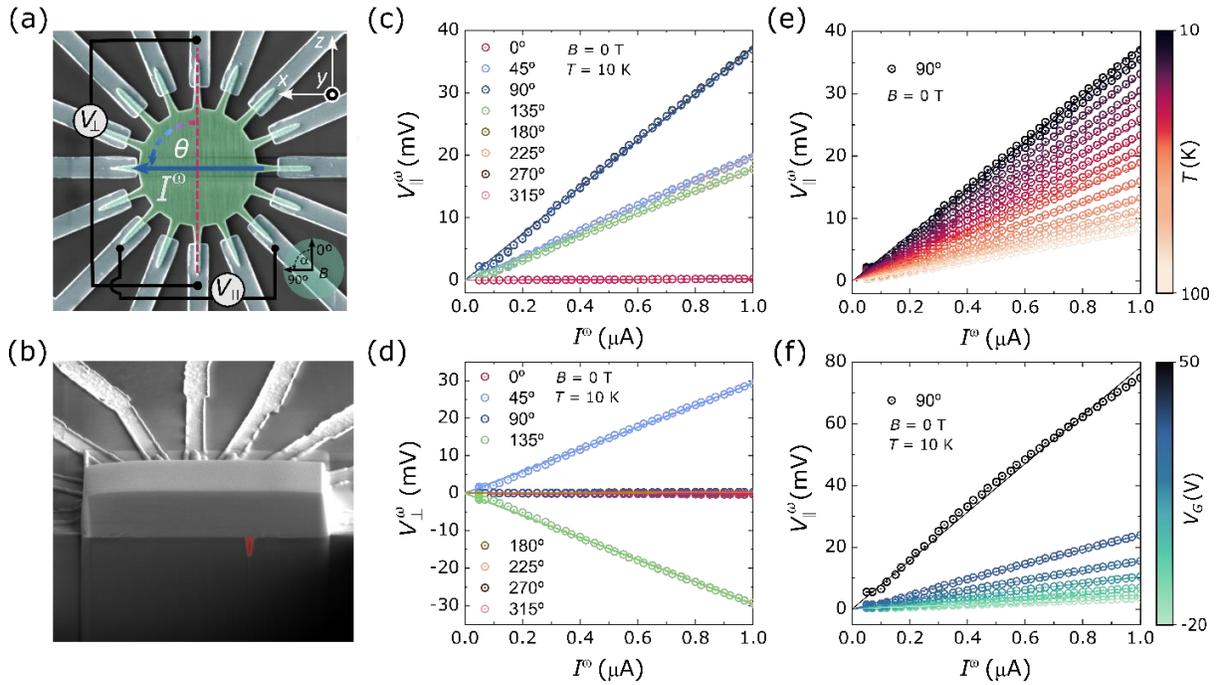

FIG. S1. Additional data from device S1. (a) False-color SEM image of device S1. The etched Te flake (green) is contacted by Pt contacts (blue). $I^\omega$ is injected through one of the 8 electrodes at angle $\theta$ from the *z*-axis, and the longitudinal ($V_\parallel$) and transverse ($V_\perp$) voltages are measured in a rotating reference frame. The angle $\alpha$ defining the orientation between the magnetic field $B$ and the *z*-axis is also drawn. (b) False-color SEM image of the transverse cut to determine the orientation of the Te atomic triangles pattern in the *x-y* plane [Fig. 3(a)]. A small notch (red) is cut into the lamellae to detect any flip of the sample. (c),(d) First-order parallel $V_\parallel^\omega$ (c) and transverse $V_\perp^\omega$ (d) voltage as a function of $I^\omega$ at 10 K for different $\theta$ angles. To plot $R_\parallel^\omega$ in Fig. 1(c) [slope of the dependences in (c)], we applied a correction factor to consider the difference between $L_\parallel$ and $L_\perp$. (e),(f) The first-order parallel voltage $V_\parallel^\omega$ along the *x*-axis ($\theta = 90°$) as a function of $I^\omega$ for different temperatures (e) and back-gate voltages (f). For all angles, temperatures, and back-gate voltages, the first-harmonic voltage depends linearly on $I^\omega$. The lines are linear fits to the experimental data.

## Section 3. Comparison between samples.

Figure S2 shows data for devices S2-S5. All samples show a similar behavior, in agreement with Te symmetry. Those shown below are: higher resistance along *x*-axis than along chiral *z*-axis [Fig. S2(b)], non-linear conductivity described by equations (2) and (3) [Fig. S2(c)], negative magnetoresistance which is maximum when the external magnetic field is aligned with *z*-axis [Fig. S2(d)], and UMR along *z*-axis [Fig. S2(e)].

For device S3, we cut a Te lamella perpendicular to the *z*-axis, as we also did for devices S1 and S2. The UMR measurement [Fig. S2(e)] indicates that the Te flake is left-handed, therefore, we orient the atomic triangle pattern pointing to $+x$ [STEM image in Fig. S2(a)]. Following this convention, we obtained a positive sign for the $\chi_{xxx}$ component, in agreement with what we discussed in the main text. For devices S4 and S5, we do not have STEM images of a Te lamella. Hence, we cannot determine the sign of $\chi_{xxx}$.

Table S1 summarizes the dimension of each sample [Figs. S1(a) and S2(a)], and the values of $\rho_{xx}$ and $\chi_{xxx}$ obtained from the fittings. We highlight that the magnitude of $\chi_{xxx}$ increases when $\rho_{xx}$ decreases. This trend can be explained considering that the microscopic origin of the effect is extrinsic. A higher number of defects results in lower resistivity, as the number of carriers increases, but also in more scattering centers which contribute to the non-linear conductivity. Nevertheless, we must note that the output voltage also depends on $\rho_{xx}^3$ [equations (2) and (3)]. Therefore, despite the lower value of $\chi_{xxx}$ for device S5 (Table S1), the output second-harmonic voltage is high [Fig. S2(c)].

TABLE S1. Comparison between devices. The table summarizes the different parameters of the devices used in this study. We cannot determine the sign of $\chi_{xxx}$ for devices S4 and S5.

| Device | S1(10K) | S2 (10K) | S3 (50 K) | S4 (10 K) | S5 (100K) |
|---|---|---|---|---|---|
| $L_{\parallel}(\mu m)$ | 3.95 | 2.95 | 3.20 | 3.14 | 1.92 |
| $L_{\perp}(\mu m)$ | 6.30 | 3.84 | 5.47 | 3.64 | 3.59 |
| $w(\mu m)$ | 0.32 | 0.63 | 0.66 | 0.83 | 0.58 |
| $t(nm)$ | 112 | 215 | 94 | ~100 | ~100 |
| $\rho_{xx}$ ($10^{-4}\,\Omega\,m$) | 3.33 | 3.08 | 11.1 | 6.96 | 56.8 |
| $\chi_{xxx}^{\parallel}(10^{-3}\Omega^{-1}V^{-1})$ | 2.13 ± 0.18 | -5.04 ± 0.86 | 0.266 ± 0.092 | 1.00 ± 0.14 | 0.041 ± 0.011 |
| $\chi_{xxx}^{\perp}(10^{-3}\Omega^{-1}V^{-1})$ | 2.49 ± 0.18 | -5.5 ± 1.6 | 0.39 ± 0.26 | 0.82 ± 0.23 | 0.060 ± 0.027 |

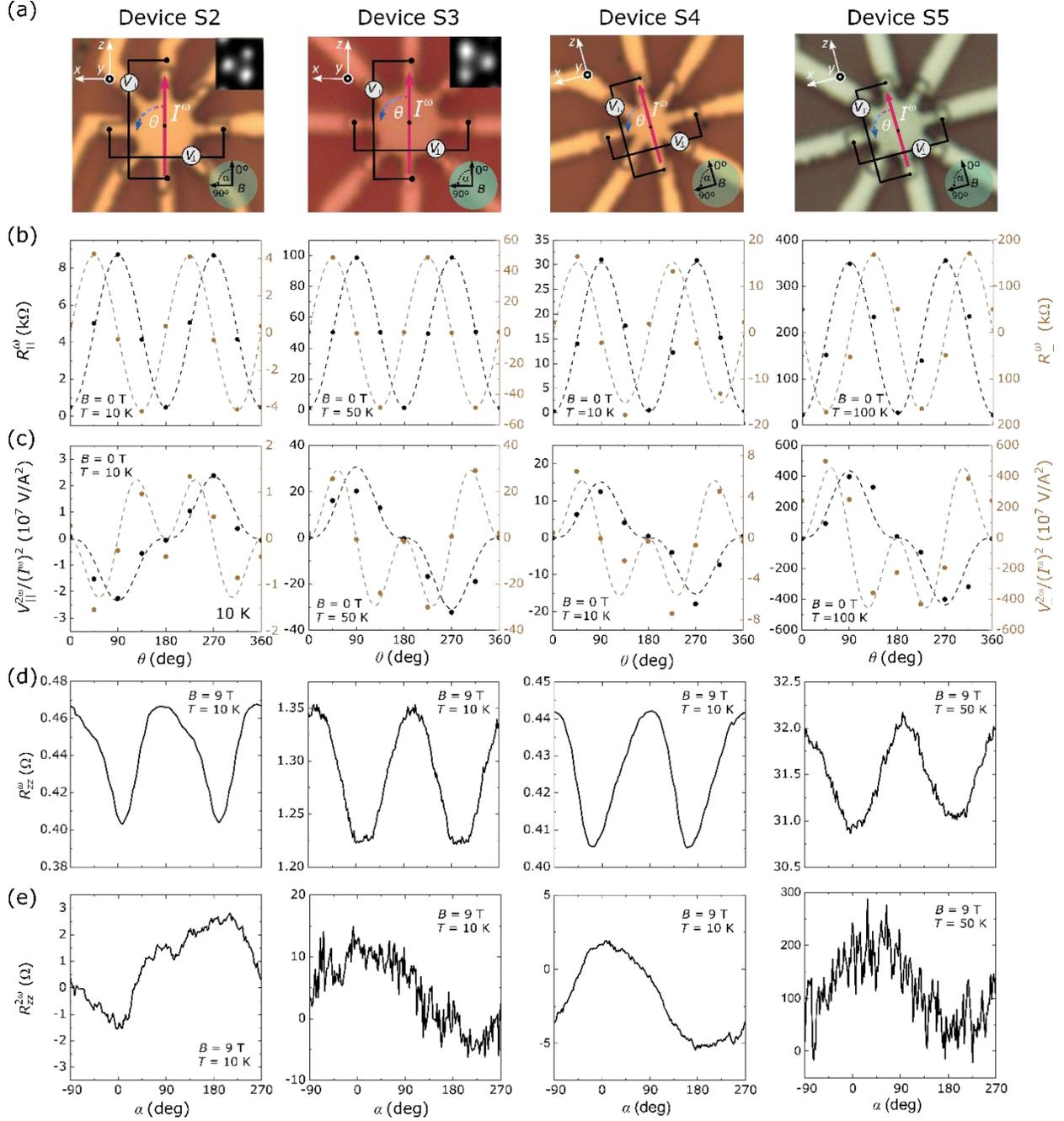

FIG. S2. Comparison between devices. a, Optical images of different devices used in this study (S2, S3, S4, and S5). Insets: STEM images of Te lamellas obtained cutting perpendicular to the $z$-axis devices S2 and S3 (Fig. S1b). (b) Parallel ($R_{\parallel}^{\omega}$) and transverse ($R_{\perp}^{\omega}$) first-harmonic a.c. resistance as a function of $\theta$-angle. The dashed lines are fits to the equations introduced in the main text. (c) The non-linear conductivity as a function of $\theta$. The dashed lines are fits to equations (2) and (3). (d),(e) $\alpha$-angle dependence of (d) $R_{zz}^{\omega}$ and (e) $R_{zz}^{2\omega}$ at 9 T. The black curve on panel (c) and panel (e) for device S2 correspond to Fig. 3(e) and Fig. 3(c), respectively.

## Section 4. Angular dependence of the second-order non-linear response

The second-order non-linear current density in response to an electric field $E$ of frequency $\omega$ can be expressed through the material's second-order non-linear susceptibility $\chi$ as $j_i^{2\omega} = \chi_{ijk} E_j^\omega E_k^\omega$. For Te with P3$_1$21 or P3$_2$21 space group symmetry, the non-linear susceptibility has 5 (2 independent) non-zero elements as shown below [48]:

$$\chi = \begin{pmatrix} \chi_{xxx} & \chi_{xyy} & 0 & \chi_{xyz} & 0 & 0 \\ 0 & 0 & 0 & 0 & \chi_{yxz} & \chi_{yxy} \\ 0 & 0 & 0 & 0 & 0 & 0 \end{pmatrix} \quad (1)$$

where $\chi_{xxx} = -\chi_{xyy} = -\chi_{yxy}$ and $\chi_{xyz} = -\chi_{yxz}$. Coordinate $z$ corresponds to the chiral $c$-axis which is the preferential direction of growth and lies in-plane. Moreover, coordinate $x$ corresponds to the $a$-axis and the cross-section STEM images indicate that lies also in-plane [Fig. 3(a)]. Therefore, for an in-plane electric field $\boldsymbol{E}^\omega = (E_x^\omega, 0, E_z^\omega)$, the non-linear current density $\boldsymbol{j}^{2\omega}$ is giving by:

$$\boldsymbol{j}^{2\omega} = \begin{pmatrix} \chi_{xxx}(E_x^\omega)^2 \\ -2\chi_{yxz} E_x^\omega E_z^\omega \\ 0 \end{pmatrix} \quad (S1)$$

Now, by using Ohm's law: $\boldsymbol{E}^{2\omega} = \boldsymbol{\rho} \cdot \boldsymbol{j}^{2\omega}$, where $\boldsymbol{\rho} = \begin{pmatrix} \rho_{xx} & 0 & 0 \\ 0 & \rho_{yy} & 0 \\ 0 & 0 & \rho_{zz} \end{pmatrix}$, we obtain:

$$\boldsymbol{E}^{2\omega} = \begin{pmatrix} \rho_{xx} \chi_{xxx}(E_x^\omega)^2 \\ -2\rho_{yy} \chi_{yxz} E_x^\omega E_z^\omega \\ 0 \end{pmatrix} \quad (S2)$$

Considering that in our experiments we measure the voltage in-plane, we can neglect the out-of-plane component ($y$-component). Therefore, we obtain the expression for the second-order in-plane electric field:

$$\boldsymbol{E}^{2\omega} = \begin{pmatrix} \rho_{xx} \chi_{xxx}(E_x^\omega)^2 \\ 0 \end{pmatrix} \quad (S3)$$

For an in-plane current density $\boldsymbol{j}^\omega = j^\omega(sin(\theta), cos(\theta))$ of amplitude $j^\omega$ at angle $\theta$ from the chiral axis ($z$-axis), the first-order electric field from Ohm's law is $\boldsymbol{E}^\omega = j^\omega(\rho_{xx} sin(\theta), \rho_{zz} cos(\theta))$. Substituting the last expression in equation (S3), we obtain the relation between the second-order in-plane electric field and the applied current:

$$\boldsymbol{E}^{2\omega} = \begin{pmatrix} (j^\omega)^2 \rho_{xx}^3 \chi_{xxx} sin^2(\theta) \\ 0 \end{pmatrix} \quad (S4)$$

Dividing in parallel and transverse components:

$$\frac{E_\parallel^{2\omega}}{(j^\omega)^2} = \rho_{xx}^3 \chi_{xxx} sin^3(\theta) \quad (S5)$$

$$\frac{E_\perp^{2\omega}}{(j^\omega)^2} = \rho_{xx}^3 \chi_{xxx} \sin^2(\theta)\cos(\theta) \tag{S6}$$

Finally, using: $E_\parallel^{2\omega} = V_\parallel^{2\omega}/L_\perp$, $E_\perp^{2\omega} = V_\perp^{2\omega}/L_\perp$, $j^\omega = I^\omega/A$, $\rho_{xx} = R_{xx}^\omega \cdot A/L_x$, we can rewrite the expression (S5) and (S6) in terms of the parameters that we directly measure in our experiments:

$$\frac{V_\parallel^{2\omega}}{(I^\omega)^2} = \frac{L_\parallel}{L_{xx}^3} A\, (R_{xx}^\omega)^3\, \chi_{xxx}\sin^3(\theta) \tag{2}$$

$$\frac{V_\perp^{2\omega}}{(I^\omega)^2} = \frac{L_\perp}{L_x^3} A\, (R_{xx}^\omega)^3\, \chi_{xxx}\sin^2(\theta)\cos(\theta) \tag{3}$$

where $A$ is the cross section of the Te flake.

Therefore, when an electric field ($E^\omega$) is applied along the x-axis, a non-linear current density is generated in the same direction [Fig. S3(a)]. Conversely, when an electric field is applied along the z-axis, the generation of a non-linear charge current is forbidden by symmetry constrains, and therefore, Ohm's law holds, $j_z = \sigma_{zz} E_z^\omega$ [Fig. S3(b)].

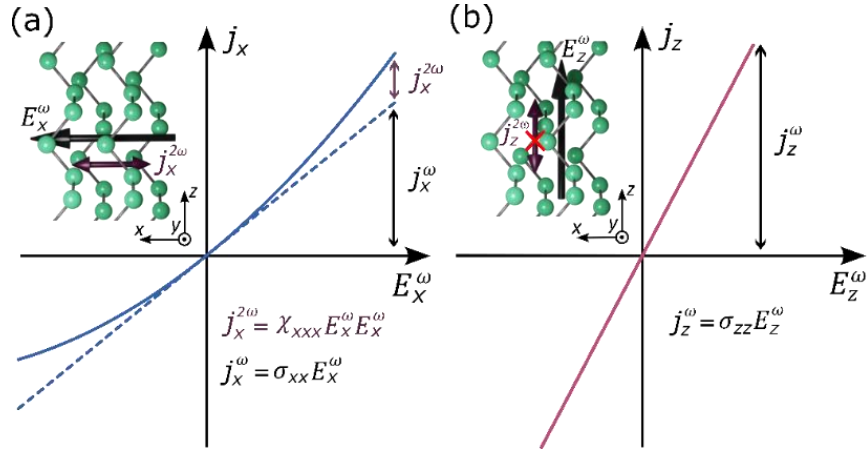

FIG. S3. Symmetry of non-linear conductivity. (a),(b) In the case an electric field $E^\omega$ is biased along (a) x-axis [(b), z-axis], the generation of a longitudinal second-order current density $j^{2\omega}$ is allowed [forbidden]. Insets: x-z plane of trigonal Te.

### Section 5. Chirality-dependent unidirectional magnetoresistance.

In this section, we will provide additional details on the unidirectional magnetoresistance response in Te which has been used to recognize the handedness of the flakes. More information can be found in Ref. [43]. Unidirectional magnetoresistance is also known as bilinear magnetoresistance or electrical magnetochiral anisotropy.

Te possesses a radial spin texture, where the spins at the Fermi level point in a direction parallel or antiparallel to $k_z$ depending on the handedness. According to the Edelstein effect, an electric field applied along the z-direction generates a current density $j_z$ and causes a redistribution of states along the $k_z$ direction, which is depicted as a $\Delta k_z$ shift of the Fermi contour. Due to the radial spin texture of Te, $\Delta k_z$ induces a homogeneous spin density with a polarization that is parallel or antiparallel to $j_z$, depending on the chirality of the flakes [FIG. S4(a),(b)]. Therefore, an electrical current in Te acquires a net spin polarization oriented along the z-axis, pointing in a direction parallel or antiparallel to $j_z$, depending on the handedness of the crystal.

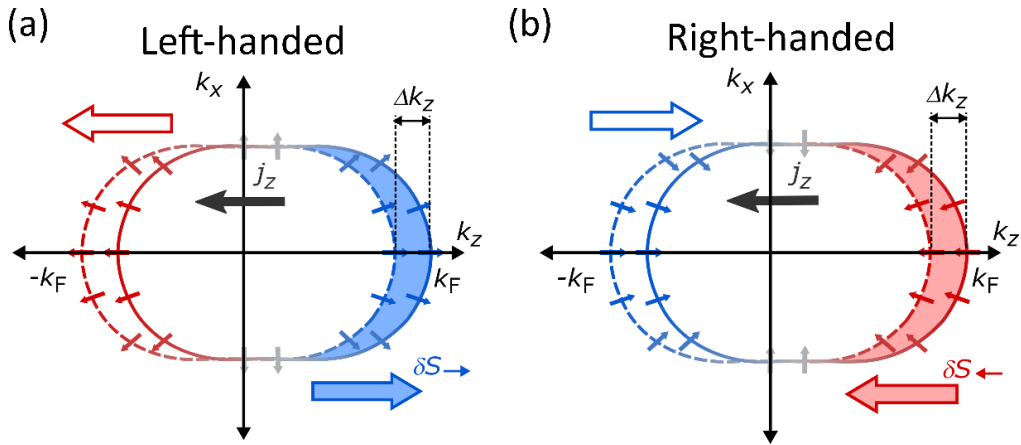

FIG. S4. Sketch representing the Edelstein effect responsible for the unidirectional magnetoresistance in Te. The shift in $k_z$ of the Fermi contours translates in the formation of spin densities oriented in opposite directions for (a) left-handed and (b) right-handed flakes. Empty big arrows mean depletion of states, filled big arrows mean higher occupation of states. Small arrows mean spins.

In these conditions, different resistance states are measured if a magnetic field is applied in a direction parallel or antiparallel to the spin polarization, giving rise to the unidirectional magnetoresistance. Consequently, the opposite spin polarization between left- and right-handed flakes allow us to recognize the crystal's handedness by magnetotransport measurements, since it introduces an opposite dependence of the resistance on the mutual orientation of current and magnetic field.

By applying currents and measuring voltages along the chiral z-axis and rotating the magnetic field in-plane as illustrated in Fig. 1(a), we analyzed the angle-dependent magnetoresistance of devices S1 and S2, measured for opposite current directions [Fig. S5(a), (b)]. For both devices, we find a strong dependence of the resistance on the relative alignment between $I_z$ and $B$, which are parallel and antiparallel at α = 0° and 180°, respectively. Device S1 is characterized by a significantly higher (lower) resistance when the current is parallel (antiparallel) to the external magnetic field. Conversely, device S2 displays lower (higher) resistance for current parallel (antiparallel) to the external magnetic field.

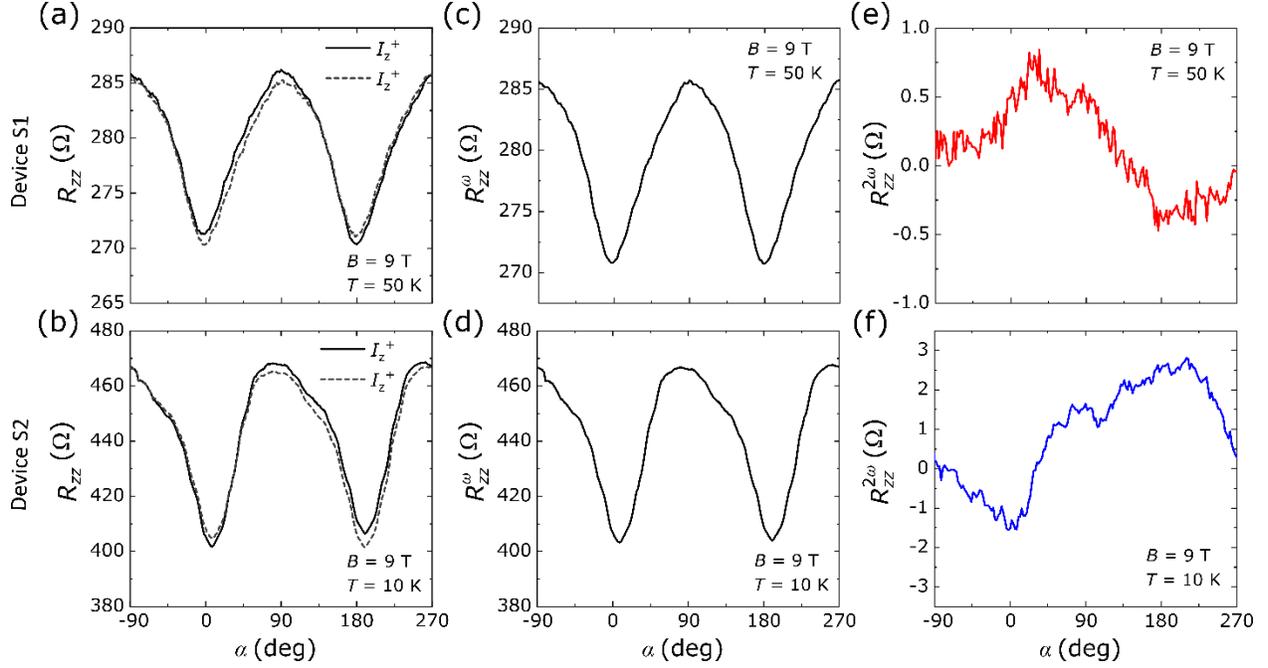

FIG. S5. Raw data of magnetoresistance measurements. (a),(b) $\alpha$-angle dependences of $R_{zz}$ measured at 9 T for device S1 at 50 K (a) and device S2 at 10 K (b). Solid and dashed lines indicate the signal obtained for positive and negative currents, respectively ($\pm I_z=\pm 1\mu A$). (c),(d),(e),(f) $\alpha$-angle dependence of (c),(d) $R_{zz}^{\omega}$ at 9 T for device S1 at 50 K (c),(e) and device S2 at 10 K (d),(f). Note that data in panel (c), (e), and (f) are the same as in Fig. 1(c), Fig. 3(b), and Fig. 3(c), respectively.

To analyze these data, we calculate the average $R_{zz}^{\omega} = [R(+I_z=+1\mu A) + R(-I_z=-1\mu A)]/2$ [Fig. S5(c),(d)] and the half difference $R_{zz}^{2\omega} = [R(+I_z=+1\mu A) - R(-I_z=-1\mu A)]/2$ [Fig. S5(e),(f)] between the resistance measured applying the current along $+z$ and $-z$-axis. These two curves are the equivalent of the first- and second-harmonic resistance in a.c. transport measurements. The average resistance, $R_{zz}^{\omega}$, has a two-fold symmetry as current direction-induced contributions are cancelled by current applied in both directions [Fig. S5(c), (d)]. The half difference resistance, $R_{zz}^{2\omega}$, presents peaks for collinear current and magnetic field (at $\alpha = 0°$ and 180°) [Fig. S5(e), (f)], corresponding to the asymmetries in Fig. S5(a), (b). This observation indicates that, in agreement with the Edelstein mechanism based on the radial spin texture of Te (Fig. S4), the spin polarization is oriented along the direction of the current, parallel to the chiral $z$-axis. Moreover, the $R_{zz}^{2\omega}$ hallmark is opposite for devices S1 and S2, demonstrating that they have opposite handedness. The sign of the $R_{zz}^{2\omega}$ was calibrated in Ref. [43] by comparing the magnetotransport measurements with STEM images which determine the handedness of the crystals. Therefore, we can claim that device S1 is left-handed and device S2 is right-handed.

## Section 6. Non-linear conductivity on device S2.

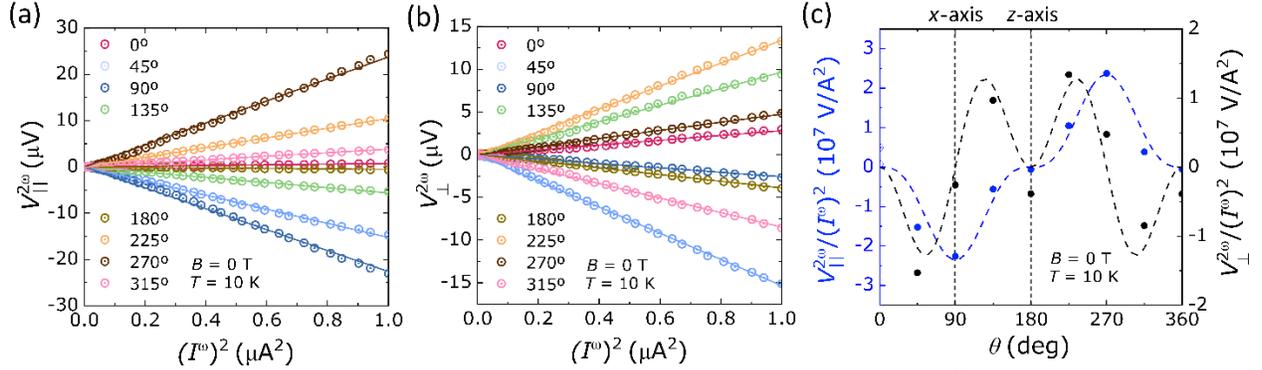

Fig. S6. Non-linear conductivity on device S2. (a),(b) Second-order parallel $V_\parallel^{2\omega}$ (a) and transverse $V_\perp^{2\omega}$ (b) voltage as a function of $(I^\omega)^2$ at different $\theta$ angles. (c) The non-linear conductivity [slopes of the dependences in (a) and (b)] as a function of $\theta$. The dashed lines are fits to the experimental data with the model described in the main text. All measurements were performed at 10 K. Note that data and fit in panel (c) are the same as in Fig. 3(e).

## Section 7. Scaling law of the non-linear conductivity.

The relation between non-linear conductivity magnitude and resistivity has been formulated by Du *et al.* [16]. This formulation is based on previous results by Hou *et al.* [50] for the mechanisms of the anomalous Hall effect. Therefore, considering the relation between $E_\perp^{2\omega}/(j^\omega)^2$ and $\chi_{xxx}$ expressed in equation (S6), the scaling law of non-linear conductivity for our experiments can be written as:

$$\frac{V_\perp^{2\omega}}{(I^\omega)^2} = C_1 \rho_{xx0} + C_2 \rho_{xx0}^2 + C_3 \rho_{xx0} \rho_{xxT} + C_4 \rho_{xxT}^2 \tag{S7}$$

with four scaling parameters:

$$\begin{aligned} C_1 &= C^{sk}, C_2 = C^{in} + C_0^{sj} + C_{00}^{sj}, \\ C_3 &= 2C^{in} + C_0^{sj} + C_1^{sj} + C_{01}^{sj}, \\ C_4 &= C^{in} + C_1^{sj} + C_{11}^{sj} \end{aligned} \tag{S8}$$

where the disorder-independent coefficients are for the intrinsic ($C^{in}$), side-jump ($C_i^{sj}, C_{ij}^{sj}$), and skew-scattering ($C^{sk}$) contributions, respectively. Since we are probing the term $\chi_{xxx}$ (diagonal), the measured effect cannot have an intrinsic origin [18], so for now on we will not consider an intrinsic contribution. The indexes $i, j$ refers to different scattering sources. We have considered one static ($i, j = 0$) and one dynamic ($i, j = 1$). $\rho_{xx0}$ is the residual resistivity due to static impurities at zero temperature and $\rho_{xxT} = \rho_{xx} - \rho_{xx0}$ is due to dynamic disorders at finite temperature. Hence, we can rewrite the expression in terms of $\rho_{xx}$, our experimental variable:

$$\frac{V_\perp^{2\omega}}{(I^\omega)^2} = \beta + \gamma \rho_{xx} + \delta \rho_{xx}^2 \tag{4}$$

where:

$$\begin{aligned}
\beta &= C_1 \rho_{xx0} + (C_2 - C_3 + C_4)\rho_{xx0}^2 = C^{sk}\rho_{xx0} + (C_{00}^{sj} - C_{01}^{sj} + C_{11}^{sj})\rho_{xx0}^2 \\
\gamma &= (C_3 - 2C_4)\rho_{xx0} = (C_0^{sj} + C_{01}^{sj} - 2C_{11}^{sj} - C_1^{sj})\rho_{xx0} \\
\delta &= C_4 = C_1^{sj} + C_{11}^{sj}
\end{aligned} \tag{S9}$$

We note that in our experimentally practical parsing of non-linear conductivity, the above definitions have not relied on identification of semiclassical processes such as side-jump scattering or skew scattering from asymmetric contributions to the semiclassical scattering rates [50]. In the full semiclassical theory, we are aware that considering the skew-scattering contribution as the sum of all the contributions arising from the asymmetric scattering rate present in the collision term of the Boltzman transport equation, there is a term which scales with $\rho_i \rho_j / \rho^3$, known as intrinsic skew-scattering [16]. However, we do not categorize this contribution as skew-scattering but rather place it under the umbrella of side-jump ($C_i^{sj}$), following the approach used in previous works studying the multivariable scaling of transport effects [50].

We studied the relation between the non-linear conductivity and resistivity in device S1, by varying the sample temperature [Fig. S7(a)] and by applying an electrostatic gate voltage [Fig. S7(b)], but also in device S2 and S3 by varying the sample temperature (Fig. S8).

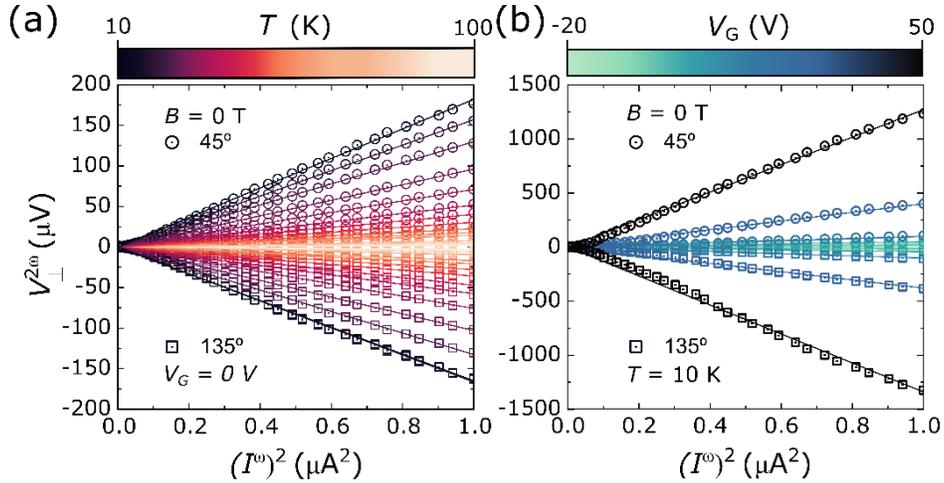

FIG. S7. Temperature and gate dependence of the non-linear conductivity. (a),(b) Second-order transverse voltage ($V_\perp^{2\omega}$) as a function of $(I^\omega)^2$ at (a) different temperatures ($V_G = 0\,V$) and (b) gate voltages ($T = 10\,K$) for $\theta = 45°$ and $135°$.

Table S2 summarizes the values for coefficients $\beta$, $\gamma$, and $\delta$ obtained fitting our experimental data to equation (4). We observe that coefficients $\gamma$ and $\beta$ dominate. Moreover, for all devices and fittings, these two terms have opposite signs. From these observations, and looking to equation (S9), we can conclude that that side-jump from dynamic sources ($C_1^{sj}, C_{11}^{sj}$) is the major mechanism that gives rise to the non-linear conductivity on Te.

The opposite sign in all terms of device S2 with respect to devices S1 and S3 is in agreement with the inversion-symmetry-odd nature of the non-linear susceptibility tensor (see Spatial inversion operation section of the main text).

TABLE S2. Scaling law of the non-linear conductivity. It shows the values of coefficients $\beta$, $\gamma$, and $\delta$ obtained from the fittings to equation (4) for devices S1, S2, and S3. In the case of device S1, we included the results for the temperature and electrostatic gate dependencies.

| Device | S1 ($T$-dep) | S1 ($V_G$-dep) | S2 ($T$-dep) | S3 ($T$-dep) |
|---|---|---|---|---|
| $\beta$ ($10^7\ V \cdot A^{-2}$) | $2.18 \pm 0.39$ | $1.03 \pm 0.31$ | $-5.04 \pm 0.99$ | $3.3 \pm 2.0$ |
| $\gamma$ ($10^{11}\ m^{-1} \cdot A^{-1}$) | $-4.44 \pm 0.54$ | $-1.78 \pm 0.54$ | $5.07 \pm 0.88$ | $-3.20 \pm 0.69$ |
| $\delta$ ($10^{15}\ m^{-2} \cdot V^{-1}$) | $2.69 \pm 0.16$ | $2.819 \pm 0.088$ | $-1.29 \pm 0.19$ | $0.595 \pm 0.058$ |

We point out that device S1 has been taking out of the cryostat between the temperature and electrostatic gate dependencies. This procedure changed the doping level, explaining the slightly different values in Table S2.

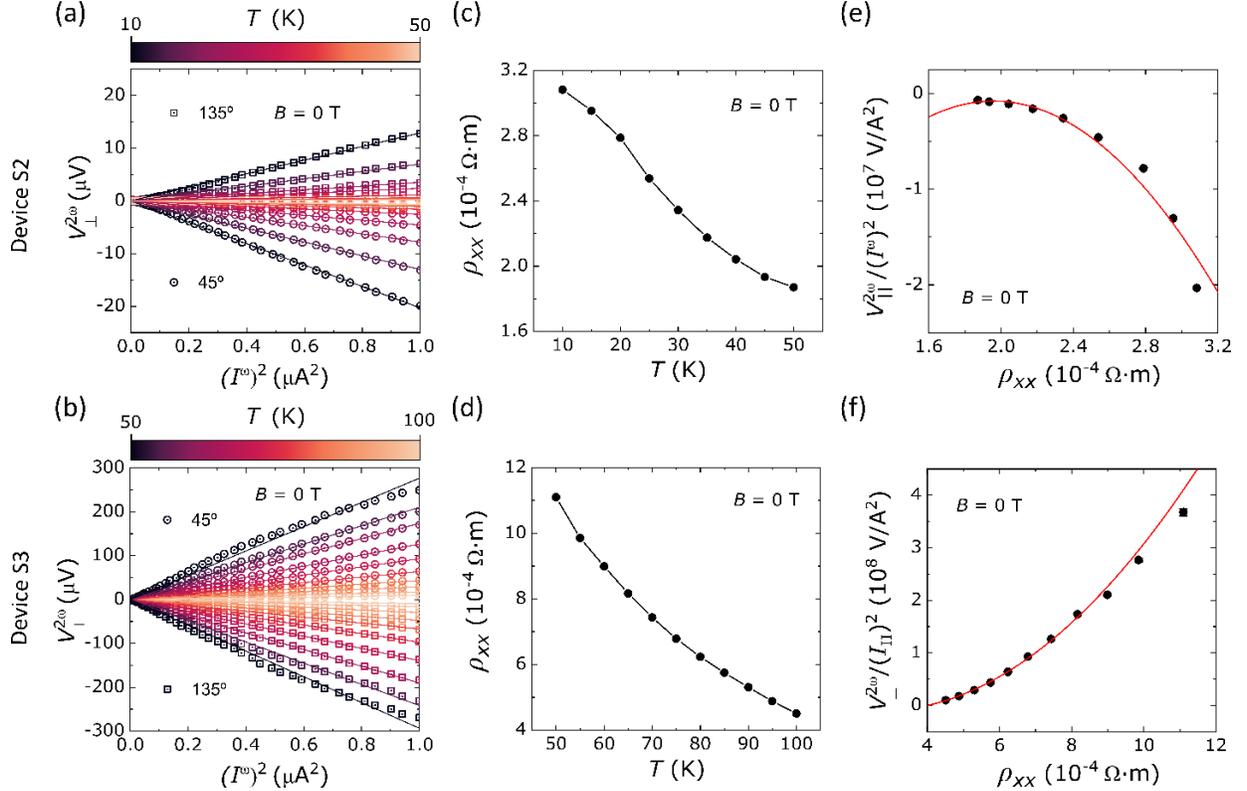

FIG. S8. Temperature dependence of the non-linear conductivity. (a),(b) Second-order transverse voltage, $V_\perp^{2\omega}$, as a function of $(I^\omega)^2$ at different temperatures for $\theta$ = 45° and 135°. (c),(d) Temperature dependence of the resistivity along x-axis ($\rho_{xx}$). (e),(f) The non-linear conductivity [taken from the slopes of $V_\perp^{2\omega}$ as a function of $(I^\omega)^2$ at $\theta$ = 45° in panels (a) and (b)] as a function of $\rho_{xx}$. The red solid lines are fits to the experimental data with equation (4). The measurements were performed in device S2: (a),(c),(e); and device S3: (b),(d),(f).

Finally, the electrostatic gate allows us to modulate the second-harmonic voltage output by a factor of 300, reaching a value of 1.2 mV for 1 µA input current [Fig. S7(b)]. Excluding engineered heterostructures, this is the highest reported value to the best of our knowledge (Fig. S9).

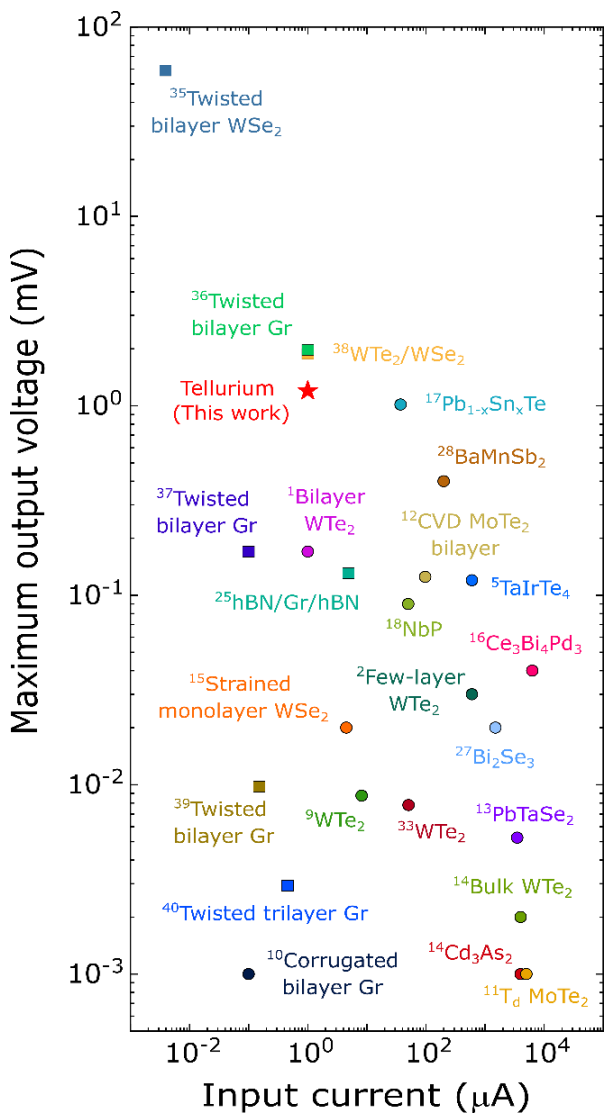

FIG. S9. Maximum non-linear output voltage versus input current values reported for materials showing non-linear transport. The engineered heterostructures appear as squares. Note that, before the name of each material, the corresponding reference is indicated.